\documentstyle[11pt,newpasp,twoside,epsf]{article} 

\begin{document} 

\title{Molecular Gas Around Young Stellar Clusters} 
\author{Naomi A. Ridge} 

\affil{FCRAO, University of Massachusetts, 619 Lederle GRC, 
Amherst, MA, 01003, USA. email: naomi@fcrao.umass.edu}
\author{Tom Megeath} 
\affil{Harvard-Smithsonian CfA, 60 Garden St., Cambridge, MA, USA.} 
\author{T. L.  Wilson} 
\affil{Steward Observatory, University of Arizona, Tucson, AZ, USA \& Max-Planck-Inst. f. Radioastronomie, Bonn, Germany.}

\vspace*{-3mm}
\begin{abstract}
We have begun a survey of the molecular gas surrounding 31 young
clusters in order to investigate the link between environment and the
resulting cluster.  We present here a preliminary comparison of two
clusters in our sample: GGD12-15 and Mon R2.  Since both clusters are
located in the MonR2 molecular cloud at a distance of 830 pc
(Carpenter 2000), observational biases due to differing sensitivities
and angular resolutions are avoided.
\end{abstract}

\vspace*{-9mm}
\section{Introduction}
Surveys for young stars in molecular clouds have shown that most stars
do not form in isolation, but in aggregates ranging from the widely
dispersed groups of stars found in the Taurus cloud, to the clusters
found in Orion, and ultimately to the ``superclusters'' such as
NGC\,3603. There is an enormous range in the richness of these stellar
groups/clusters, from several stars in Taurus to more than a thousand
stars in NGC\,3603, all within regions of a few cubic parsecs.  To
understand this enormous range in richness and the dependence of the
richness on the molecular gas environment, we have undertaken a
comparative study of both the molecular and stellar components of a
sample of 31 clusters within 1 kpc of the Sun.

\vspace*{-5mm}
\section{Observations}
We have obtained maps of the molecular gas surrounding the two young
stellar clusters in $^{13}$CO 1--0 and C$^{18}$O 1--0 at 45$''$
resolution with the FCRAO 14m telescope. These observations were
followed up with higher resolution (30$''$) C$^{18}$O 2--1 On-the-Fly
(OTF) maps, obtained at the Heinrich Hertz Telescope (HHT), operated by
the Submillimeter Telescope Observatory (SMTO).

Figure 1 shows contours of the C$^{18}$O 2--1 emission overlaid on a
2MASS K-band image. The dense gas traced by the C$^{18}$O
seems to trace the ring structure apparent in the stellar component of
Mon\,R2.  The gas around GGD\,12-15 appears elongated in the east-west
direction.  
\begin{figure}[h]
\plottwo{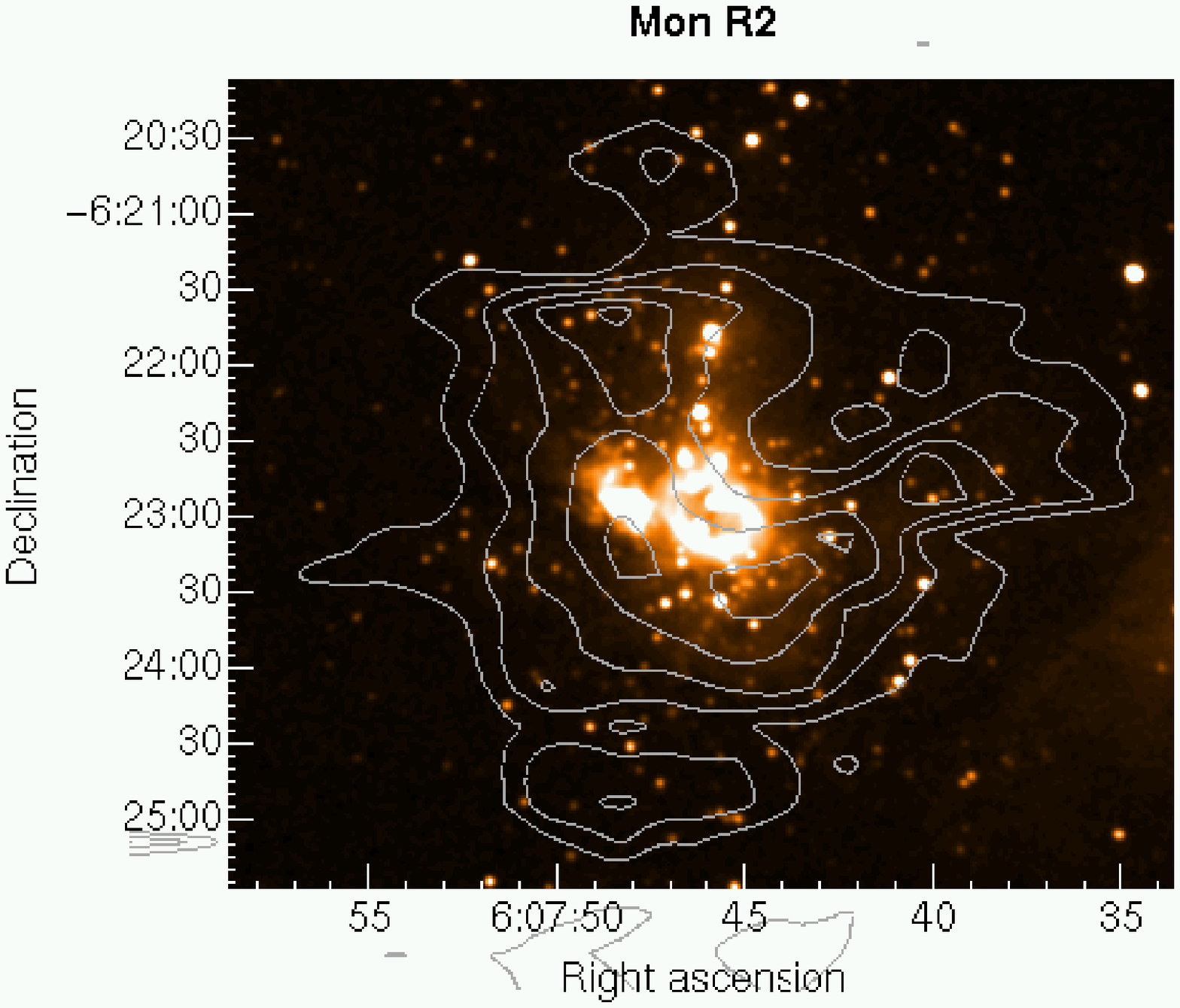}{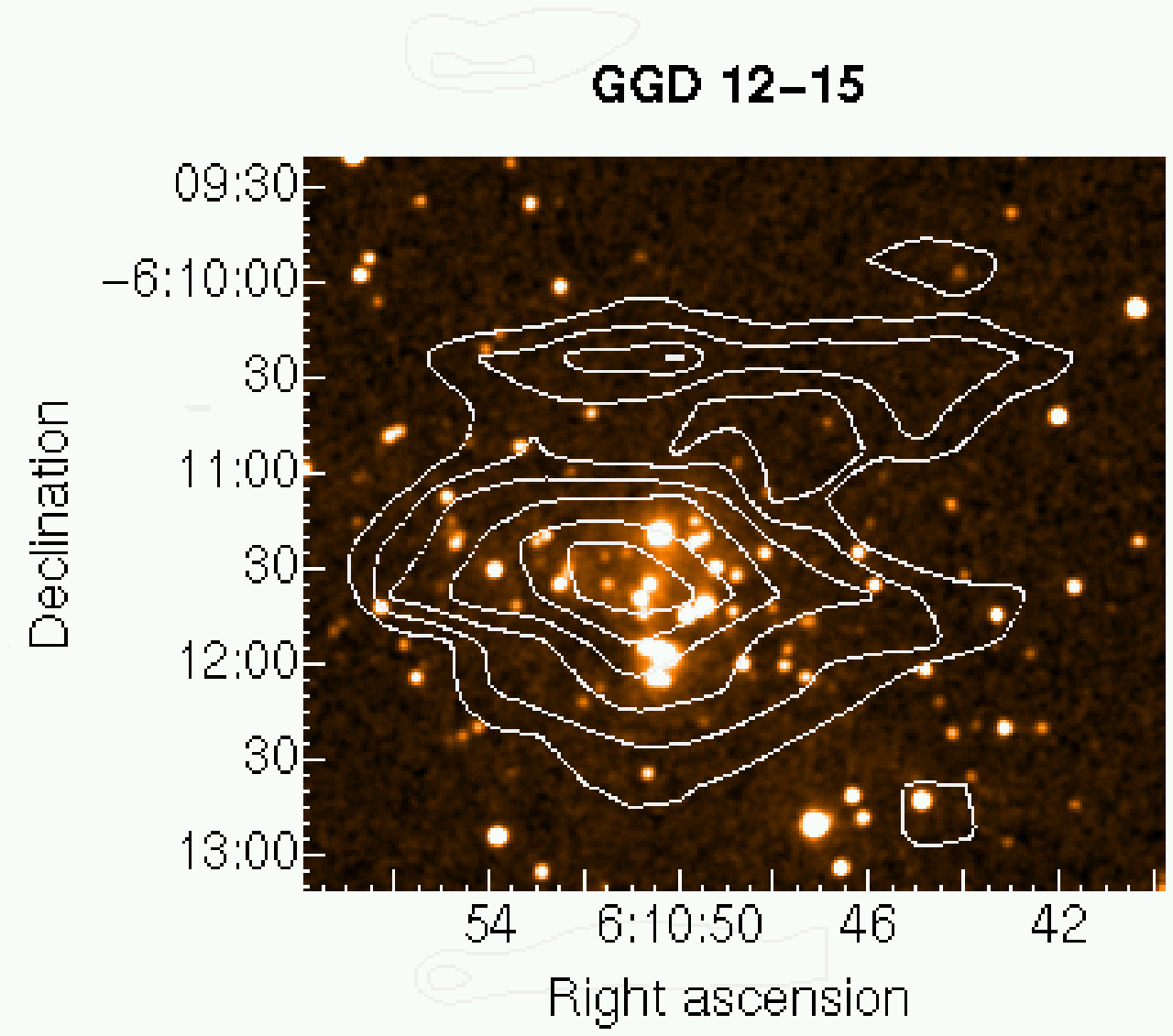}
\vspace*{-2mm}
\caption{C$^{18}$O 2--1 contours of integrated antenna temperature
overlaid on a 2MASS K-band image.  Left: Mon R2. Base contour is 7K,
contour interval is 2K.  Right: GGD\,12-15. Base contour is 7K,
contour interval is 2K.}
\end{figure}
The table below compares the gas and stellar properties of the two
clusters. The stellar properties are taken from Carpenter (2000),
while the gas properties are derived from our C$^{18}$O observations.
\begin{table}
\centering
\begin{tabular}{lccc}
\hline
Cluster & N$_{\rm stars}$ & M$_{\rm gas}$(total) & mean $\Delta$V\\
        &               & x 10$^4$\,M$_{\sun}$ & km\,s$^{-1}$\\
\hline
Mon\,R2 & 371&9.2 &1.68\\
GGD\,12--15 & 134 &5.4 & 2.26\\
\hline
\end{tabular}
\end{table}
Figure 2 shows maps of the C$^{18}$O line width, centered on the same
position as fig.\ 1 . The ring-like morphology seen in the K-band
image of Mon\,R2 can also be clearly seen in the line-width map. The
horizontal line apparent at a Dec offset $\sim$ +100$''$ is an
artifact due to lower signal--noise ratio in this region of the
map. There is also a region of low line-width north-east of the
stellar cluster.  It is harder to pick out any structure in the
noisier line-width map of GGD\,12-15, but it does appear to increase
towards the centre.

\begin{figure}[h]
\plotfiddle{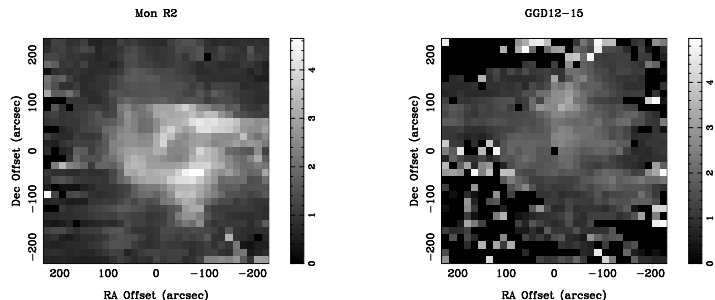}{3cm}{0}{100}{100}{-150}{-15}
\caption{Maps of the C$^{18}$O line width. Left: Mon\,R2. Right: GGD\,12-15}
\end{figure}

\end{document}